# Thermal stability and optoelectronic behavior of polyaniline/GNP (graphene nanoplatelets) nanocomposites

Pralhad Lamichhane[1], Dilli Dhakal[1], Ishan N Jayalath[2], Ganga Raj Neupane[3], Nishan Khatri[4], Parameswar Hari[3], Kaan Kalkan[4], Ranji Vaidyanathan[1].

[1]School of Materials Science and Engineering, Oklahoma State University, Tulsa, OK, USA

[2] Department of Chemistry, Oklahoma State University, Stillwater, OK, USA

[3]Department of Physics and Engineering Physics, University of Tulsa, Tulsa, OK, USA

[4] School of Mechanical and Aerospace Engineering, Oklahoma State University, Stillwater, OK

**Abstract**

Polyaniline and graphene nanoplatelets (PANI-GNP) nanocomposites are synthesized by in situ oxidative polymerization of polyaniline using an oxidizing agent, ammonium peroxy-disulphate (APS). The mass of GNP in the nanocomposites varied by 5, 10, and 15 wt% compared to PANI. The synthesized polyaniline coated graphene nanoplatelets (PANI-GNP) nanocomposites are chemically characterized and using Fourier Transform Infrared Spectroscopy (FTIR), Raman spectroscopy, Scanning electron microscopy (SEM), UV-Vis spectroscopy, and X-ray diffraction analysis (XRD). FTIR and Raman spectroscopy analysis confirmed the uniform coating of polyaniline on GNP. The SEM micrograph and XRD pattern demonstrate the polymerization quality and crystallization degree of samples. UV-Vis analysis showed a decrease in the bandgap of polyaniline, which confirms that nanocomposites are more suitable for optoelectronic application because of variation in the bandgap. TGA analysis showed the thermal stability of PANI is increased with the increased mass of GNP. This study suggests the potential of GNP as a filler for efficient modification in the morphological, electrical, optical, and thermal properties of PANI.

**Keywords:** nanocomposites, functionalization, polymerization, bandgap, optoelectronics

1. Introduction

Polyaniline is one of the most promising organic semiconducting materials among intrinsically conducting polymers. Polyaniline has several advantages over other conducting materials, such as ease of preparation, low cost, non-toxic, tunable electrical conductivity, and interesting intrinsic redox properties [1-4]. However, some limiting factors such as low thermal stability, higher band gap, and electrical instability prevent the excellence of PANI in a wide range of applications. The use of graphene-based materials as additives for PANI has been tested successfully to mitigate the above drawbacks [5]. There is a synergetic correlation between conductivity and capacitive properties of carbon allotropes and electrochemical properties of PANI. Therefore, hybrid materials of PANI/graphene and PANI/carbon nanotubes can be used in supercapacitors, energy storage, solar cells, electromagnetic shielding, electronic displays, and sensors [5, 6].

Carbon allotropes, including carbon nanotubes (CNTs) and graphene, have successfully blended with PANI using various processing methods. Examples of processing methods for the fabrication of PANI and carbon nanocomposites include solution blending, melt blending, in situ polymerization, and chemical modification processes. Almasi et al. studied the bandgap variation of PANI by adding MWCNT by in situ polymerizations of PANI [7]. Chen et al. prepared a graphene polyaniline hybrid by one-step intercalation polymerization of aniline inside the expanded graphite to study the electromagnetic properties [8]. Sheng et al. have studied the properties of graphene/PANI multilayer films used for electrochromic devices by layer-by-layer assembly [9]. Madhabi et al. studied the thermal, electrical, and dielectric properties of PANI and reduced graphene oxide (rGO) nanocomposites fabricated by in situ reductions of GO [10]. Despite having outstanding properties of this carbon nanofiller, there is still a lack of large-scale production of pure CNTs and graphene because of the high production cost. This led to demotivating the use of CNTs and graphene in industrial applications. Graphene nanoplatelets (GNPs), consisting of a few layers of graphene nanosheets having a few nanometers in thickness, are a much cheaper option than CNTs. Because of the outstanding electron transfer mechanism and several band structure layers, GNP is the most prominent material combined with conducting polymer for manufacturing electronic, optoelectronic, electrochemical, and gas sensors devices [11].

Modification of PANI by GNP forming PANI-GNP nanocomposite achieved promising performance in many applications such as sensors, actuators, microelectronics, electrochromic devices, smart fabric, supercapacitors, energy storage devises, conformal antenna systems, and electromagnetic shielding devices [12, 13]. Because of having a greater surface area, GNP shows better interaction with the polymer matrices through interfacial bonding. Due to the presence of a conjugated π-electron system in PANI and GNP, composite also possesses electrostatic interactions such as H-bonding or π - π interactions between GNP nanosheets and PANI. There is mutual compatibility of PANI and GNP for the properties enhancement mechanism.

In this study, the GNP surface was modified with PANI by in situ polymerization of aniline. The synergetic effect between PANI-GNP was evaluated for the overall electrical, physical, optical, and thermal properties enhancement mechanism. The PANI-GNP hybrid material was chemically characterized using Fourier Transform Infrared Spectroscopy (FTIR), Raman spectroscopy, x-ray diffraction (XRD), Scanning electron microscopy (SEM), and UV-Vis spectroscopy. The thermal stability of the material was evaluated using thermogravimetric analysis (TGA).

## 2. Experimental procedure

### 2.1. Materials

Graphene nanoplatelets (average surface area 120 - 150 $m^2/g$ and average particle diameter of 5 μm (Grade M particles)) were purchased from XG Science. Sulfuric acid ($H_2SO_4$), hydrochloric acid (HCl), aniline (99.5 % pure), and ammonium persulfate (APS) were purchased from Sigma-Aldrich. All the chemicals and reagents were used without further modification or purification.

### 2.2. Synthesis of PANI-GNP nanocomposites

The PANI-GNP nanocomposites were synthesized by in situ anionic polymerization of aniline in the presence of GNP. In a typical procedure, different weight percentages of GNP (5, 10, 15 wt% ) were dispersed in 100 mL of 1M HCl using an ultrasonicator and magnetic stirrer. After 20 min, 1.92 g of aniline was added into GNP dispersion with continuous stirring. The pre-cooled APS solution (4.56 g of APS dissolved in 100 mL of 1M HCl) was added dropwise into the reaction mixture. The reaction was carried out under constant stirring for 4 h in an ice bath; ice bath was used to increase the polymerization reaction rate by keeping the temperature about 0 –5 °C. After 4 h, the resulting dark green mixture was kept in the refrigerator overnight for complete polymerization of aniline. The PANI-GNP composites were filtered and washed with deionized water, HCl, and acetone two times each to remove unreacted aniline. Finally, the product was dried in an oven at 80 °C for 10 h to obtain a PANI-coated GNP nanocomposite. The weight ratio of PANI in composites was measured for a different set of samples in the final product of PANI-GNP nanocomposites. The pure PANI was also prepared with the same procedure adding no GNP to the solution. The probable reaction mechanism of cationic polymerization of aniline in the presence of HCl and APS is shown in Fig. 1.

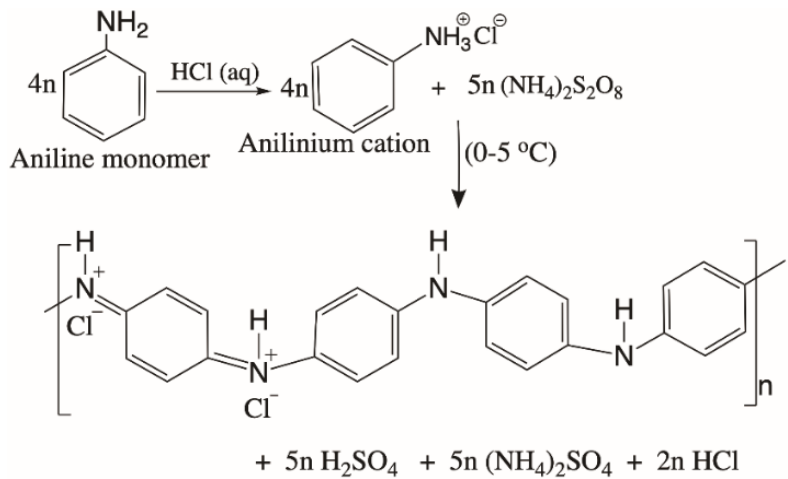

**Fig. 1.** Oxidative polymerization of aniline in the presence of APS

*2.3. Characterization*

Fourier-transform infrared spectroscopy (FTIR) was used to investigate the chemical interaction between PANI and GNP. FTIR spectrum was recorded using Nicolet iS10 FTIR spectrometer with 64 scans and 4.0 cm$^{-1}$ at room temperature.

Raman spectroscopy was observed by WITec alpha 300R Raman microscope with a 532 nm laser, spot size 5 μm. The microscope has a 20x objective lens of 0.40 numerical aperture, 600 lines/mm grating, and 100 μm confocal aperture.

XRD spectra were collected using a Bruker AXS D8 Discover X-ray Diffractometer with a general area detector diffraction system (GADDS) Vantec 500 2D detector. Fixed wavelength (1.54056 Å) was used for the experiment scanning angle of 2θ from 10° to 40°.

The surface morphology of the GNP, PANI, and PANI-GNP nanocomposites was investigated using an FEI Quanta 600 field-emission gun environmental scanning electron microscope (FESEM).

UV-Vis spectroscopy analysis was used to study the optical properties of all samples. The absorption spectra were collected at the wavelength range of 200 – 800 nm at room temperature.

Thermogravimetric analysis of PANI, pristine GNP, and PANI-GNP nanocomposites was performed using a high-resolution TGA analyzer (model Q-50, TA Instruments). During the analysis, all the samples were heated at a rate of 20 °C/min from room temperature to 800 °C under 50mL/min of continuous airflow.

## 3. Results and discussion

### 3.1. Fourier transform infrared spectroscopy (FTIR)

The surface functional groups of GNP, PANI, and PANI-GNP have been verified by using FTIR spectroscopy, and each spectrum is shown in Fig. 2a. FTIR spectra of pristine GNP exhibited the bands at 3445 cm$^{-1}$, 1635 cm$^{-1}$, 1265 cm$^{-1}$, 1097 cm$^{-1}$, 802 cm$^{-1}$ indicating the stretching vibrations of O–H bond, C=C stretching vibrations, C–OH stretching vibrations, C–O stretching vibrations, and C–H bend vibration, respectively [14, 15]. The presence of –OH, –C=C, –C–OH, and –C–O groups on the GNP surfaces make the strong interaction between GNP and PANI.

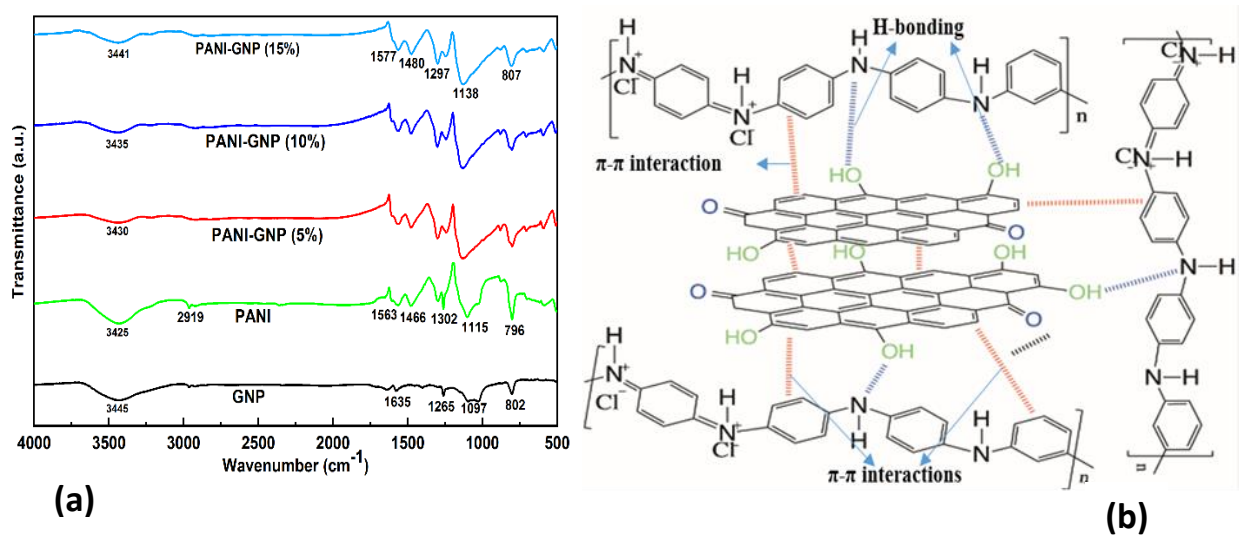

**Fig. 2.** (a) FTIR spectra of PANI, GNP, and PANI-GNP nanocomposites (b) the schematic representation of possible interactions between graphene sheets and PANI molecules

In the PANI spectrum, the signals at 3425 cm$^{-1}$, 1563 cm$^{-1}$, and 1466 cm$^{-1}$ are ascribed to the N-H stretching vibrations, C=C stretching vibration for the quinoid ring, and benzenoid ring, respectively [15, 16]. The peaks at 2919 cm$^{-1}$, 1302 cm$^{-1}$, 1115 cm$^{-1}$, and 796 cm$^{-1}$ represent C–H bending vibration, aromatic amine C–N stretching vibration, in-plane aromatic C–H bending vibration, and out-of-plane aromatic C–H bending vibration, respectively [15, 17]. The peaks of PANI-coated GNP nanocomposites were slightly shifted (measured at 3441, 1577, 1480, 1297, 1138, and 807 cm$^{-1}$ for PANI-GNP (15%)), showing the interactions between PANI and GNP. The bands of C=C stretching vibrations in PANI shifted to higher wavenumber in nanocomposites demonstrating the π-π interaction and formation of hydrogen bonding between GNP and PANI [8, 18]. The schematic of the interaction between the PANI and GNP is shown in Fig. 2b.

### *3.2. Raman Spectroscopy*

Raman spectroscopy provides a molecular fingerprint to characterize carbon-containing materials. Raman spectra of GNP, PANI, and PANI-GNP nanocomposites are shown in Fig. 3. GNP spectrum shows three major Raman-active peaks at 1345 cm$^{-1}$, 1582 cm$^{-1}$, and 2717 cm$^{-1}$ representing D-band, G-band, and 2D-band, respectively. D-band corresponds to structural defects and disorders of carbon atoms from sp$^2$ to sp$^3$ hybridization, and sharp G-band is assigned to vibration of sp$^2$ hybridized carbon atoms in the basal plane [15, 19]. The 2D bands correspond to the number of stacked graphene layers in nanoplatelets [20].

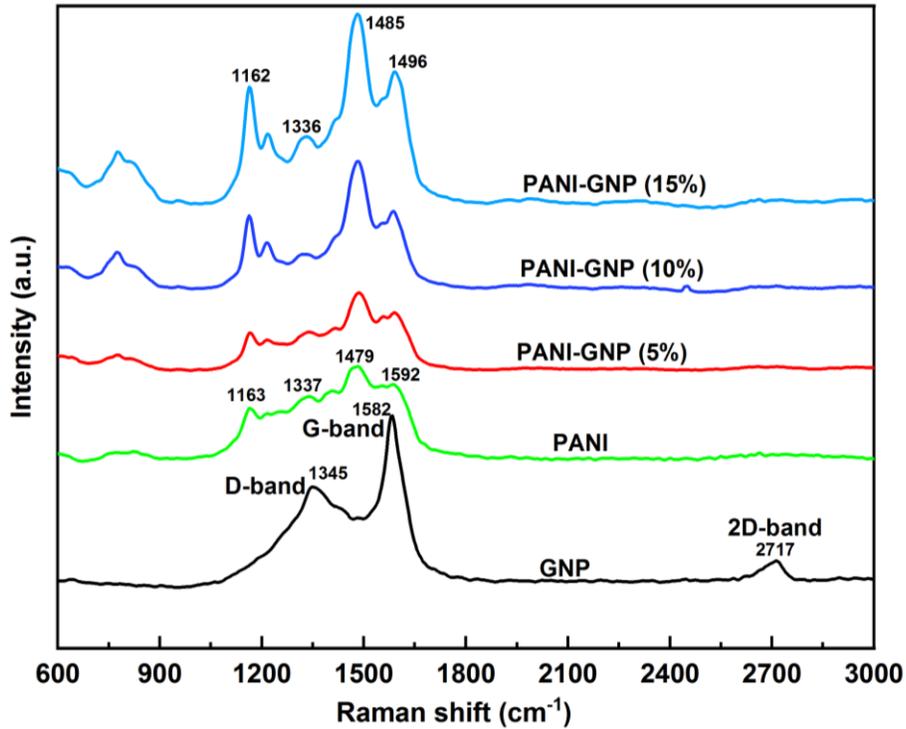

**Fig. 3.** Raman spectra of Pure GNP, PANI, and PANI-GNP nanocomposites

For PANI, C-H bending deformation in the benzenoid ring at 1163 cm$^{-1}$, C-N$^+$ stretching vibration in bipolaron structure at 1337 cm$^{-1}$, C=C stretching vibration at 1479 cm$^{-1}$, and C=C stretching of quinoid at 1592 cm$^{-1}$ were observed [21, 22]. All the spectra of PANI-GNP nanocomposites are similar to pure PANI showing no spectra related to D and 2D bands. Since Raman spectroscopy gives information on the surface of the materials, most PANI peaks were observed on PANI-GNP samples with slight shifts in wavelengths. This infers that PANI has been successfully coated on the GNP surface. It is also worth noting that the changes on the peaks in PANI-GNP nanocomposites show the formation of some interactions between PANI and GNP.

### *3.3. X-ray Diffraction (XRD) analysis*

The X-ray diffraction patterns provide the crystallinity information of the materials. The XRD spectra of PANI and PANI- GNP nanocomposites are shown in Fig. 4. For pure PANI, the major characteristics peaks observed (2θs) at 15.03°, 20.31°, 22.46°, 25.34°, and 34.15° are attributed to diffraction planes (011), (020), (021), (200), and (310), respectively. These peaks are comparable to standard PANI-HCl emeraldine salt, as reported in previous works [23, 24]. Polyaniline is a two-phase system as it contains the benzenoid and quinoid groups. In the crystalline phase, the chains are parallel and ordered in a close-packed array, while in the amorphous regions, the chains are mobile and disordered [25]. Based on the diffraction patterns, the synthesized polyaniline has a more ordered structure compared to graphene and has semi-crystalline nature.

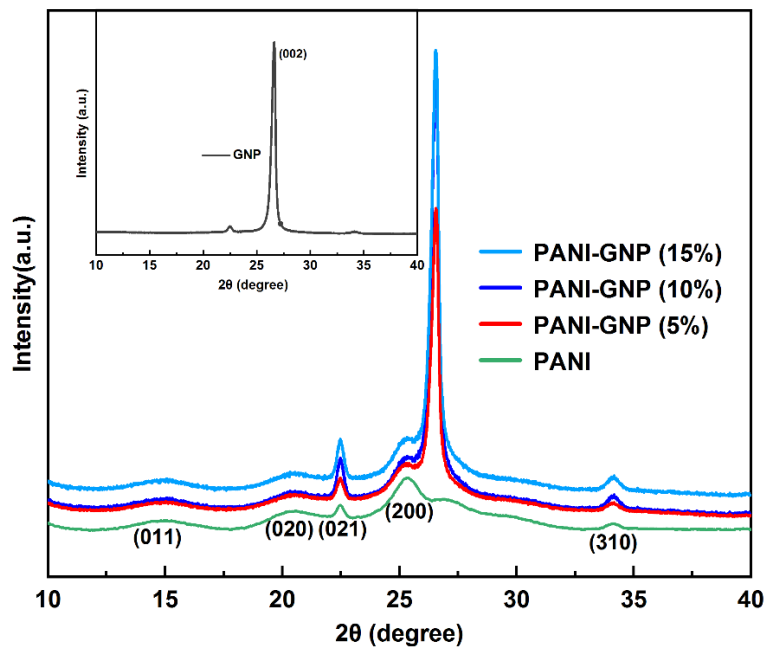

**Fig.4.** XRD pattern of PANI, GNP, and PANI-GNP nanocomposites

Pristine GNP showed a single high intense characteristic peak at 2θ of 26.5°, corresponding to (002) crystal plane of graphite nanoplatelets [26]. The PANI-GNP nanocomposites demonstrate the hybrid of diffraction peaks of both PANI and GNP. This result confirms the existence of both PANI and GNP during the in situ polymerization of aniline on the GNP surface. It was observed that the intensity of diffraction peak for the (002) plane increased with the higher loading of GNP, showing a more graphenic character in nanocomposites.

### *3.4. Scanning Electron Microscopy (SEM) analysis*

The SEM images of pristine GNP, PANI, and PANI-GNP are demonstrated in Fig. 5. GNP was observed as flaky and translucent having a relatively smooth surface with small folds without any other substances adhered to the surface. The morphology of PANI showed an irregular shape consisting of granular circular-shaped structures. The micro-granular structure formed by the aggregation of small globular structures is attributed to intermolecular interactions. However, the presence of GNP significantly disturbed the morphology of PANI globular. All PANI-GNP nanocomposites showed a rougher surface compared to that of pristine GNP (Fig. 5a), indicating the presence of PANI. The small, thickly accumulated, and rough substances on the GNP show that PANI is coated on the GNP surface through in-situ polymerization.

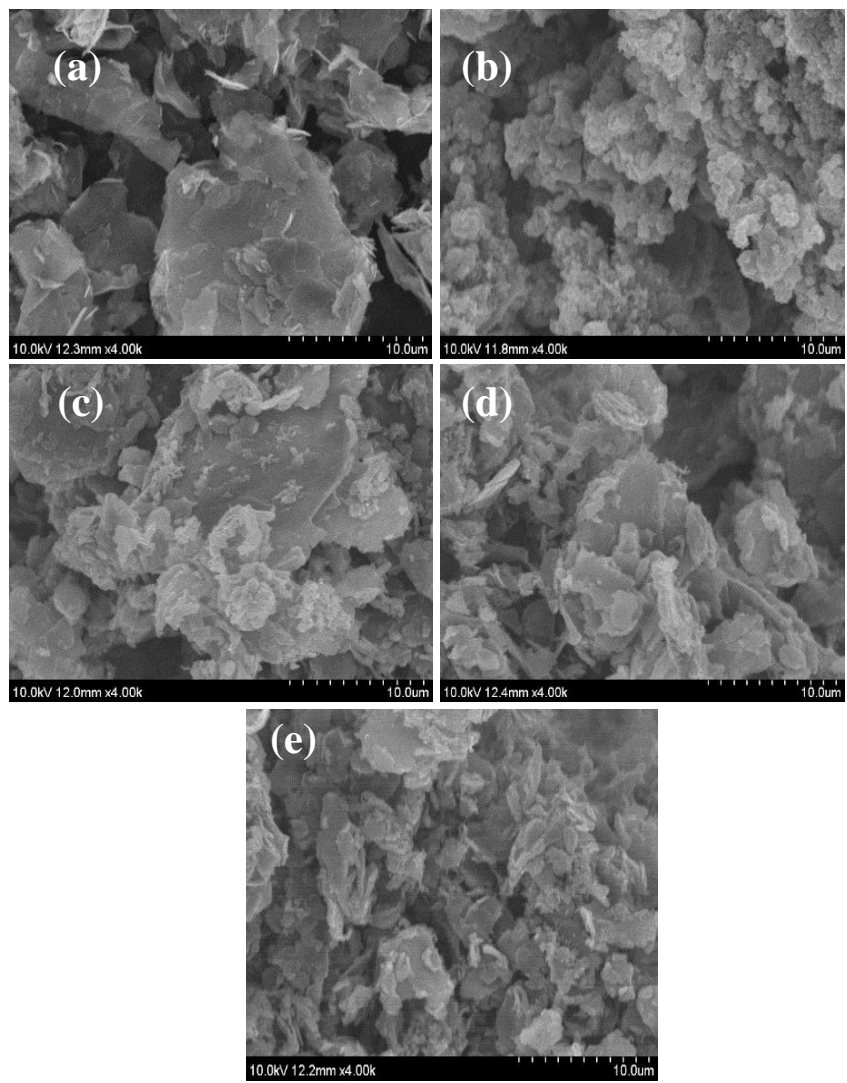

**Fig. 5.** SEM images of (a) pristine GNP, (b) PANI, (c) PANI-GNP (5%), (d) PANI-GNP (10%), and (e) PANI-GNP (15%)

### 3.5. UV-Vis spectroscopy

UV-visible spectroscopy of PANI, pristine GNP, and PANI- coated GNP nanocomposites are shown in Fig. 6. All samples showed the characteristic absorption peaks at 360 – 400 nm and 440 – 490 nm wavelengths. The band at 360 – 400 nm is associated with the π-π* transition of a benzenoid ring of polyaniline. The absorption band at 440 – 490 nm is due to the formation of polaron-π and polaron-π* transition [27-29]. Shifting of absorption bands towards the higher wavelength region on the GNP added composites indicates the graphene doping on the quinonoid ring of polyaniline to form charge transfer in PANI-GNP composites.

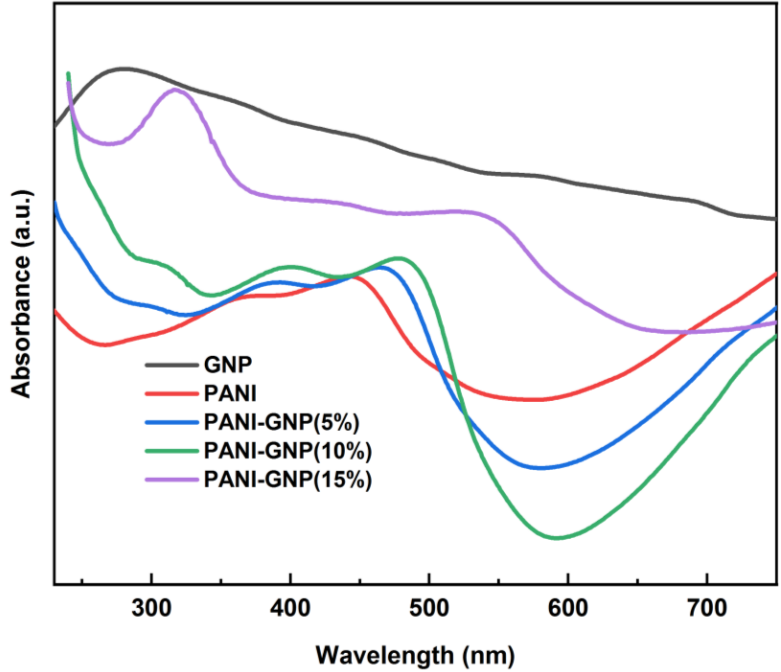

**Fig. 6.** UV-Visible absorbance spectra of PANI, pristine GNP, and PANI-GNP composites

### *3.6. Optical bandgap calculation*

The bandgap energy of all the samples was calculated using a Tauc plot relation as shown in equation 1, which is a plot of energy (hν) against the square of absorption (αhν)² [30].

$$(\alpha h\nu)^2 = A(h\nu - E_g) \qquad (1)$$

where A is the edge width parameter, ν is the frequency of the incident photon, $E_g$ is the bandgap energy, and α is the absorption coefficient. Energy (hν) was plotted against the square of absorption (αhν)². The bandgap energies were calculated by extrapolating a straight line over the slope of the main peak where the absorption is near zero, as shown in Fig. 7 (a–e). The calculated bandgap energy of pristine GNP, PANI, and PANI-GNP composites are presented in table 1. The bandgap energy of pristine GNP was measured as 2.59 eV as shown in Fig. 7a. Increasing the concentration of GNP on the composites reduced the bandgap of PANI, showing 4.26 eV for pure PANI and 3.65 eV for PANI-GNP (15%) nanocomposites. GNP addition creates new excitation energy levels by charge transfer from GNP to PANI and reduces the bandgap [7, 31]. Sharma et al. also observed similar results while adding MWCNT on PANI nanocomposites [5]. They observed a decrease in bandgap from 4.17 eV to 3.55 eV when 15 mg MWCNT was added to the composites. Hence it can be observed that the bandgap is controlled by varying the ratio of the materials in composites, and one can obtain the desired value by selecting the proper composition ratio.

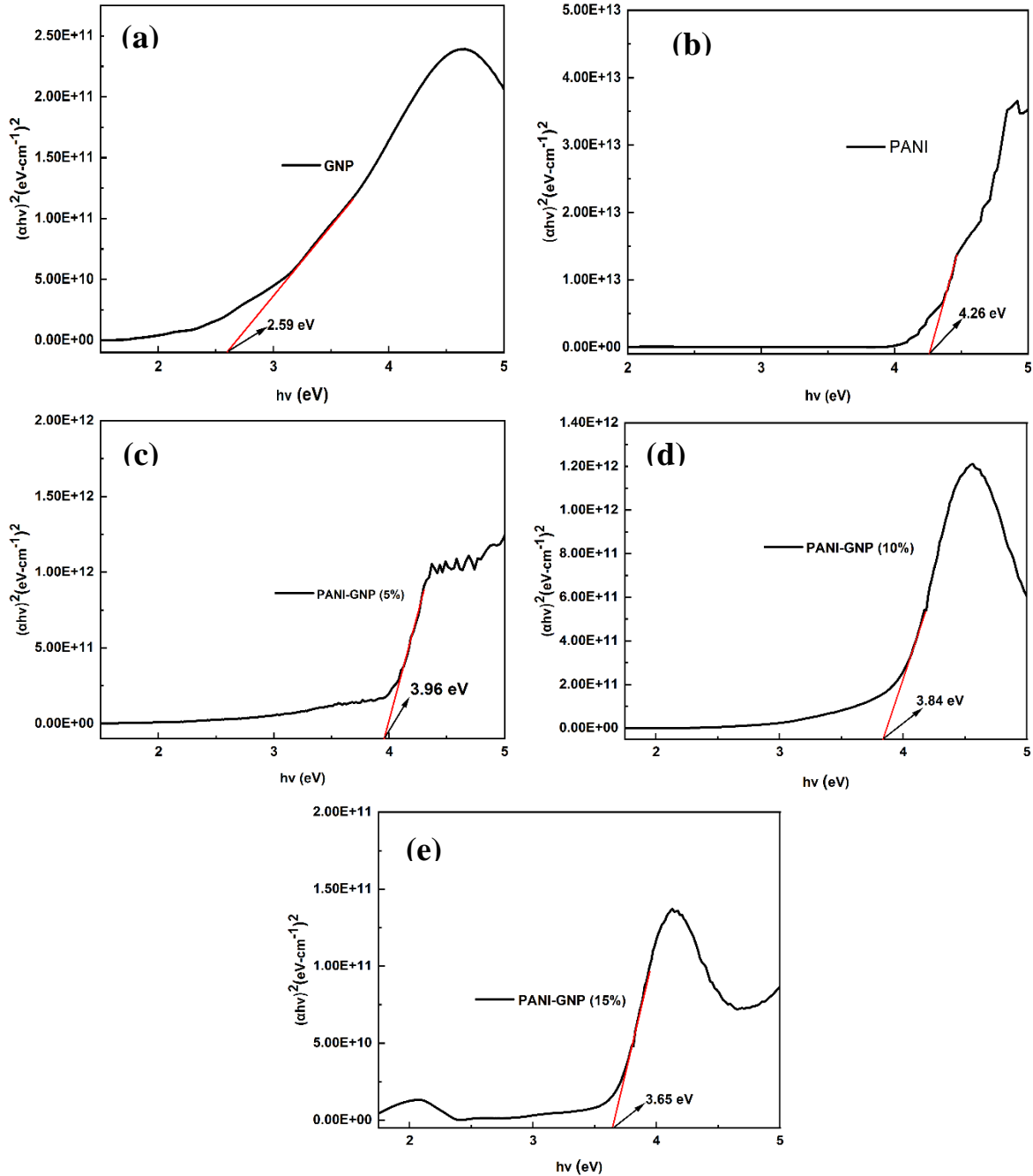

**Fig. 7.** Bandgap calculation of pristine GNP, PANI, and PANI-GNP nanocomposites

**Table 1.** Bandgap energy of samples

| Samples | GNP | PANI | PANI-GNP (5%) | PANI-GNP (10%) | PANI-GNP (15%) |
|---|---|---|---|---|---|
| Bandgap (eV) | 2.59 | 4.26 | 3.96 | 3.84 | 3.65 |

*3.7. Thermogravimatric analysis (TGA)*

Thermal stability is an important parameter in determining the manufacturing and applications of materials under different environmental conditions. The thermal stability of polymeric materials highly depends on the process parameters and their loading states [32, 33]. TGA/DTG (differential thermo-gravimetric) analysis was performed to study the thermal stability of PANI and PANI-GNP nanocomposites and shown in Fig. 8. TGA/DTG curves show two-step weight loss of PANI, GNP, and PANI-GNP nanocomposites. The first weight loss of PANI up to 225°C is due to the vaporization of water molecules and other impurities. Another major weight loss occurs in the temperature range of 300 – 625 °C due to the breaking of the polymer chain. Similarly, for all PANI-GNP nanocomposites, the first-step weight loss up to 225 °C is attributed to the elimination of some volatile species and moisture adsorbed on the surface of GNP. The second-step weight loss in the range of 300 – 625 °C is due to the thermal degradation of hexagonal carbon in GNP and the breaking of the PANI polymer chain. The thermal stability of pure PANI is less than other nanocomposites (~2% char yield), while GNP has the highest thermal stability throughout the temperature range. The thermal stability was increased with increasing the GNP concentration in the nanocomposites; the char yield was ~38% with GNP in PANI. This improvement in the thermal stability of PANI is associated with the formation of a non-flammable network of GNP serving as a fire shield in a polymer matrix [34, 35]. Furthermore, the addition of GNP would constrain the mobility of PANI chains via physical or chemical interfacial bonds so that decomposition temperature increases which may account for higher thermal stability of nanocomposites compared to pure PANI. The presence of such a shield inhibits the rate of heat transfer and polymer degradation [36].

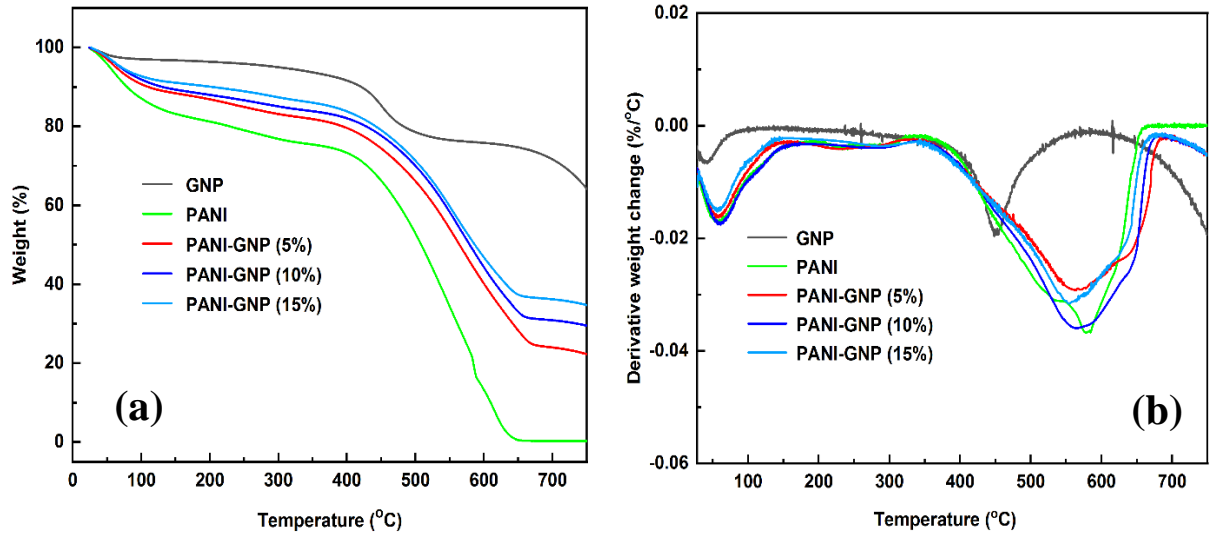

**Fig. 8.** Thermal stability of PANI and PANI-GNP nanocomposites (a) TGA, and (b) DTG

## 4. Conclusions

Polyaniline-coated graphene nanoplatelets were successfully fabricated by in situ polymerizations of aniline. The chemical characterization and surface analysis of PANI–GNP composites were confirmed by FTIR, Raman spectroscopy, and SEM analysis. The occurrence of (002) and (021) XRD peaks verified the existence of GNP and PANI in PANI - GNP nanocomposites. The more graphenic character of the nanocomposites was observed by XRD with the higher loading of GNP in nanocomposites. The UV- Visible studies showed a decrease in the optical bandgap of PANI, confirming the interaction between GNP and PANI. It is observed that the addition of a higher volume of GNP decreases the bandgap of PANI from 4.26 to 3.65 eV. Similarly, TGA results showed an increase in char yield from ~2% to ~ 38% inferring the enhancement in thermal stability of polyaniline due to the addition of GNP.

**Acknowledgments**

This work was financially supported by the National Science Foundation I-Corps site [project number 1548003]; and Helmerich family endowment funds through the Varnadow Chair funds to Dr. Ranji Vaidyanathan.